# Active Learning Exploration of Transition Metal Complexes to Discover Method-Insensitive and Synthetically Accessible Chromophores

Chenru Duan[1,2], Aditya Nandy[1,2], Gianmarco Terrones[1], David W. Kastner[1,3], and Heather J. Kulik[1,2]

[1]Department of Chemical Engineering, Massachusetts Institute of Technology, Cambridge, MA 02139, USA

[2]Department of Chemistry, Massachusetts Institute of Technology, Cambridge, MA 02139, USA

[3]Department of Biological Engineering, Massachusetts Institute of Technology, Cambridge, MA 02139, USA

Transition metal chromophores with earth-abundant transition metals are an important design target for their applications in lighting and non-toxic bioimaging, but their design is challenged by the scarcity of complexes that simultaneously have optimal target absorption energies in the visible region as well as well-defined ground states. Machine learning (ML) accelerated discovery could overcome such challenges by enabling screening of a larger space, but is limited by the fidelity of the data used in ML model training, which is typically from a single approximate density functional. To address this limitation, we search for consensus in predictions among 23 density functional approximations across multiple rungs of "Jacob's ladder." To accelerate the discovery of complexes with absorption energies in the visible region while minimizing MR character, we use 2D efficient global optimization to sample candidate low-spin chromophores from multi-million complex spaces. Despite the scarcity (i.e., ~ 0.01%) of potential chromophores in this large chemical space, we identify candidates with high likelihood (i.e., > 10%) of computational validation as the ML models improve during active learning, representing a 1,000-fold acceleration in discovery. Absorption spectra of promising chromophores from time-dependent density functional theory verify that 2/3 of candidates have the desired excited state properties. The observation that constituent ligands from our leads have demonstrated interesting optical properties in the literature exemplifies the effectiveness of our construction of a realistic design space and active learning approach.

## Introduction

Transition metal chromophores are an important design target because they play an important role in many chemical and biological processes ranging from natural light harvesting[1-3], light-emitting technologies[4] to photocatalysis[5,6]. Due to the delicate interplay[7,8] required to tune complex properties, it is challenging to use a standard Edisonian approach[9] to simultaneously alter metal–ligand interactions, ligand field strength, electron donating/withdrawing effects, and the relative energetic positioning between the ground and excited state potential energy surfaces. A notable exception is a recent work[10] that utilized high-throughput experiments to identify heteroleptic Ir(III)-based chromophores. Nevertheless, to facilitate scalable materials design, transition metal chromophores made with 3*d* earth-abundant metals with $d^6$ electron configuration are preferred relative to their state-of-the-art 4*d* and 5*d* metal (e.g. Ru(II) and Ir(III)) analogs[7,8].

The combination of virtual high-throughput screening (VHTS)[11-19] and machine learning (ML)[20-26] shows great promise and has started to address combinatorial challenges in accelerating the design of functional molecules and materials. In this approach, a large set of materials or molecules are studied with density functional theory (DFT) to develop structure–property relationships.[17,27-33] Then either supervised learning (i.e., forward) models[20-23,34] are trained to screen a large pre-constructed design space or generative (i.e., inverse) models[35,36] are applied to obtain candidate molecules with targeted properties. A train-then-predict approach usually requires too much computational time for data generation and can be sensitive to how the compounds are selected for training. Therefore, active learning with Bayesian optimization[37-39] has been recognized as an attractive paradigm for balancing data acquisition in ML model training (i.e., exploration) and ML-model-based prediction (i.e., exploitation) for chemical discovery[40-43], demonstrating a 500-fold acceleration[44] compared to random search.

Despite the success of this active learning approach in many applications, there remain significant challenges that prevent experimental realization of the predictions yielded by computational workflows. First, the DFT outcome depends on the density functional approximation (DFA) choice. A DFA that works well on certain systems may fail prominently on other systems due to the approximations made in the exchange-correlation functional[14,45,46]. When a single-DFA approach is used in VHTS, the DFA choice can lead to large biases in the data sets generated, which in turn biases the candidates the ML models recommend[47]. For transition metal complexes (TMCs) in particular, the electronic structure of a TMC is sometimes dominated by static correlation[48] that would make DFT error-prone, and predictions can be highly sensitive to DFA choice. Additionally, it is difficult to guarantee that the predicted lead molecules are synthesizable, despite the ability to add explicit constraints to ML models[49-51]. For TMCS the synthesizability problem becomes multiplicative[52] (i.e., all ligands comprising a TMC need to be synthesizable and compatible with complex formation).

In this work, we apply efficient global optimization[53] (EGO) to discover $3d^6$ Fe(II)/Co(III) transition metal chromophores in a design space with 32.5M TMCs. We address the outstanding challenge of synthesizability of candidate chromophores by carefully crafting the design space with constraints using synthetically accessible fragments and ligand symmetries in Cambridge Structural Database (CSD). We avoid bias from DFA choice by applying a DFA-consensus approach[47] that considers property evaluation as an ensemble of predictions from 23 DFAs that span multiple rungs of "Jacob's ladder"[54]. Our active learning approach successfully identifies promising transition metal chromophores and is estimated to exhibit a 1,000-fold acceleration compared to the random sampling. We reveal that Co(III) complexes with large, strong-field ligands with more saturated bonds are preferred as candidate transition metal chromophores. By invoking Hammett tuning effects[55] and introducing functionalization on compounds, we further enriched the number of potential transition metal chromophores and verified our most promising candidates with time-dependent DFT (TDDFT) calculations.

## Results and Discussion

**Design space.** We construct and explore a hypothetical design space of TMCs where all the constituent fragments (i.e., metal ions and ligands) are synthetically accessible (Figure 1). We further constrain the TMCs in the space to contain three bidentate ligands (e.g. $Fe(II)(bpy)_3$) and restrict ourselves to $d^6$ Fe(II) or Co(III) metal centers, based on the precedent of this coordination type supporting efficient chromophores. We limit the number of unique ligands in a complex to two in order to promote the likelihood of synthesizability. We started with 5,173 CSD ligands that we curated in previous work[56], including by assigning the charge states of the ligands. From this set, we selected bidentate ligands that contain common elements (i.e., H, B, C, N, O, F, Si, P, S, Cl, Br, and I) and ≤ 25 heavy (i.e., non-hydrogen) atoms, leaving a set of 812 ligands (Supporting Information). Combined with either Fe(II) or Co(III), the constraint of forming a complex with up to two unique bidentate ligands in an octahedral complex with three bidentate ligands produces 2×812=1,624 homoleptic and 2×812×811=1,317,064 heteroleptic TMCs. We refer to these 1.3M TMCs as the "base complexes". Hammett tuning is a commonly adopted strategy in experiments to fine tune the electronic properties of a complex by adding electron-donating or electron-withdrawing functional groups on conjugated rings. Here, we consider three distinct functionalization positions and ten functional groups, expanding the final design space to 32.5M functionalized TMCs, which we refer to as the "functionalized complex space" (Figure 1, see details in *Exploring the functionalized design space*).

**Active learning procedure and design criteria.** In transition metal chromophores, the photo-excited state should have a lifetime that is sufficiently long, such that the resulting chemical potential can be redirected before it is lost to unproductive competing pathways. Correspondingly, it is advantageous to have the photo-excited electron populate a long-lived metal–ligand charge transfer (MLCT) state and avoid low-lying metal-centered (MC) states that deactivate electron transfer from the expected photo-excited state. Therefore, we target complexes with low-spin (LS) ground states to increase the likelihood of MLCT states and to destabilize MC states[8]. We also desire target complexes to have weak multi-reference (MR) character. Avoiding high MR character has the benefit of avoiding complexes for which even a consensus DFT approach is likely to be inaccurate. It is possible to efficiently estimate MR character from fractional occupation number DFT as the contribution from nondynamical correlation[57,58] (i.e., $r_{\mathrm{ND}}$[59], see *Methods*). When this value is low, we also anticipate a lack of deleterious low-lying electronic states. In addition, the absorption energy should fall within the wavelengths of the visible spectrum, ranging from 1.5 eV (825 nm) and 3.5 eV (350 nm). The absorption energy is estimated from Δ-SCF calculations[60], which are more robust to DFA choice than the HOMO–LUMO gap from orbital energies (see *Methods*).

We use EGO with a 2D probability of improvement (P[I]) criterion to sample TMCs with LS ground states in a target zone of [1.5 eV, 3.5 eV] for Δ-SCF gap and [0, 0.307] for $r_{\mathrm{ND}}$ as candidate transition metal chromophores (Figure 1). The 2D P[I] is used to estimate the overall model probability (i.e., total area from the prediction and its model uncertainty) of residing within the target region and lacks the term present in some active learning algorithms that favor data points with higher uncertainty beyond a defined Pareto front. At odds with Pareto front-focused optimization, the 2D P[I] score employed here is amenable to the design goal of discovering a range of equally valid Δ-SCF gaps with modest $r_{ND}$ values. The cutoff of 0.307 for $r_{ND}$ was chosen based on our previous work[61] as a distinguishing cutoff for TMCs with weak vs. strong MR character. For EGO, we largely followed the established protocols from our previous work[43,44]: complexes in each new generation were selected by *k*-medoids sampling over the full design space. We then used DFT to evaluate the Δ-SCF gap, $r_{ND}$, and ground spin state of these complexes. After combining the new data with data from previous generations, we retrained our ML models. Lastly, we used the updated ML models to evaluate the ground spin state and 2D P[I] of the complexes. We selected the top 2,000 TMCs for *k*-medoids sampling to obtain the 200 complexes for DFT simulation in the next generation. Importantly, because both the ground state assignment and Δ-SCF gap are sensitive to DFA choice, we adopted a DFA consensus procedure[47], where we considered an ensemble of 23 DFAs that cover the broad spectrum of "Jacob's ladder" of functionals to increase the robustness of our lead candidate chromophores (Supporting Information Figures S1–S3). Specifically, we only retained complexes when a majority of DFAs (i.e., 70% or > 16 of 23) predict the complex to have a LS ground state (Supporting Information Table S1). During the evaluation of 2D P[I], we consider 23 ML models separately trained on each DFA, from which the Δ-SCF gap and its corresponding model uncertainty (i.e., from a calibrated distance in latent space[62]) are estimated. The $r_{ND}$ is computed only from an ML model trained on a single DFA (i.e., B3LYP) because trends in $r_{ND}$ values have been shown to be insensitive to the functional once calibrated[63]. On the contrary, the Δ-SCF gap is averaged over the 23 models due to its relatively high DFA sensitivity (Supporting Information Figure S3). The resulting 23 2D P[I] values derived from the $r_{ND}$ and Δ-SCF gap values are averaged to rank and sample TMCs in the next generation.

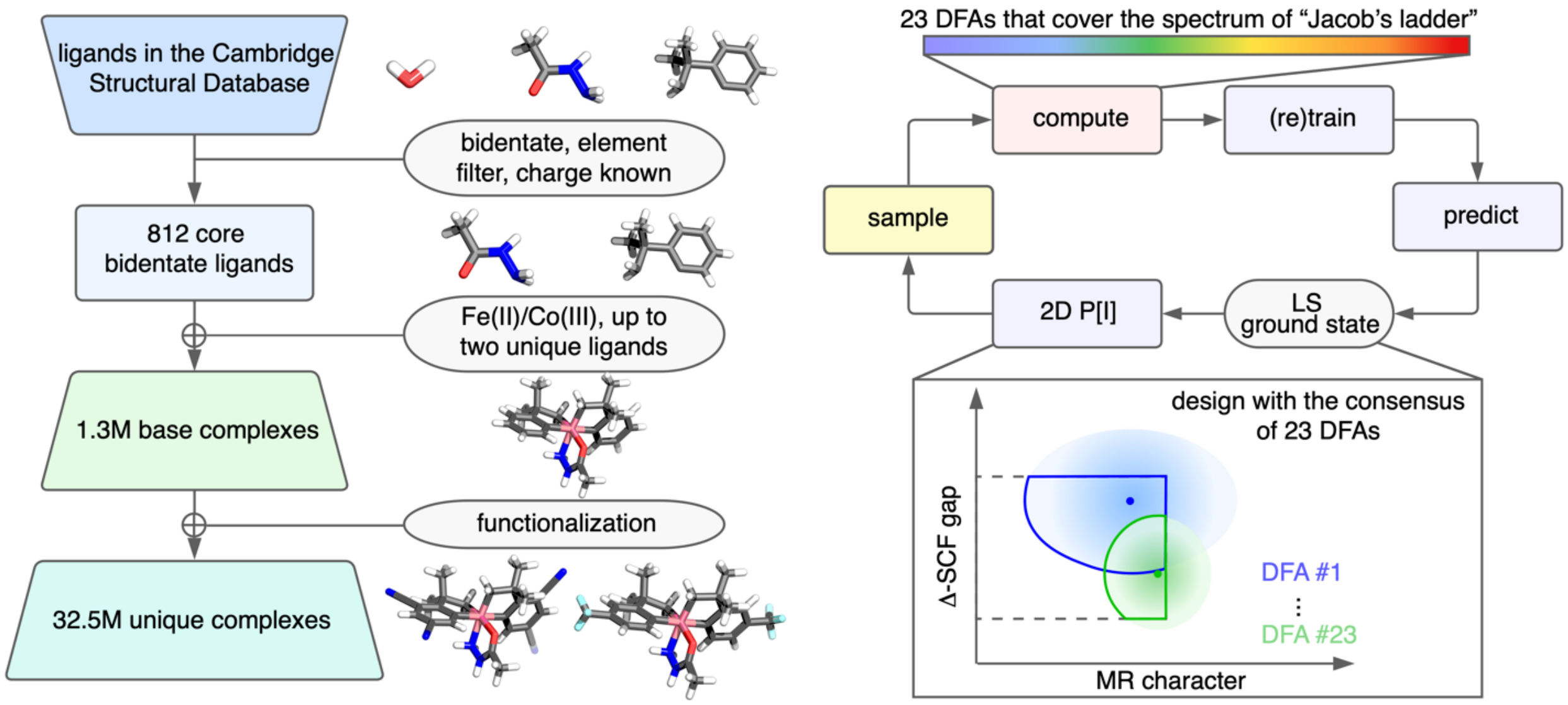

**Figure 1.** (left) Hierarchical assembly of the 32.5M complex design space for transition metal chromophores. All ligands in the CSD are first filtered to retain bidentate ligands with common elements and ≤ 25 heavy atoms, and known charge. The resulting 812 bidentate ligands are paired with either Fe(II) or Co(III), under the constraint that each complex has ≤ 2 unique ligand types, to form a design space of 1.3M base complexes. Lastly, these complexes are expanded to the full design space of 32.5M complexes with functionalization on the coordinating rings with a series of electron-donating or electron-withdrawing functional groups. (right) Active learning for discovering DFA-consensus-designed transition metal chromophores. DFT simulations are performed with 23 DFAs that span multiple rungs of "Jacob's ladder", which are used to iteratively train ML models. These ML models are applied to predict the ground spin state and Δ-SCF gap of complexes for 23 DFAs, and the MR character and their corresponding uncertainties. These quantities are used to select complexes with LS ground states and to evaluate the 2D P[I] of the design space to sample candidate complexes to compute in the next generation. The inset is an illustration of the ML prediction (solid dot), uncertainty (shaded area), and effective 2D P[I] area (solid outline) for multiple DFAs (blue and green) with respect to a target zone (rectangle with dashed lines).

**Active learning on the 1.3M base complexes.** We observe a strong negative linear correlation between Δ-SCF gap and $r_{\mathrm{ND}}$ for the 2,000 TMCs sampled in the initial generation (gen-0), which introduces difficulties for identifying candidates with simultaneously low Δ-SCF gap and $r_{\mathrm{ND}}$ (Figure 2). This negative linear correlation exists because a small Δ-SCF gap generally suggests the existence of low-lying excited states, which would lead to high $r_{\mathrm{ND}}$ as MR character arises from near-degenerate occupied and virtual orbitals. In addition, we find that LS complexes often have stronger MR character, and thus higher $r_{ND}$ relative to their HS counterparts because they can access more configuration state functions[61] (Supporting Information Figure S4). The nature of this multiple objective search for transition metal chromophores suggests that TMCs that can fulfill all our design requirements will be scarce. Indeed, for the 2,000 TMCs in gen-0, no complex with an LS ground state matches our target criteria with desirably low Δ-SCF gap and $r_{ND}$ (Figure 2). To put this in context, the lack of compounds in gen-0 suggests an extremely low probability, $p$, of a TMC residing in the target: when $p$ is as low as 0.030%, there would still be 1/3 chance of finding at least one target complex in 2,000 random trials. Our ML models trained on gen-0 data also give a similarly conservative estimate that only 0.018% of TMCs have a 2D P[I] > 1/3, i.e., have a one-third chance of simultaneously fulfilling the two design criteria (Figure 2).

Despite the initial absence of promising transition metal chromophores, we use active learning to discover lead TMCs in the target zone. The distributions of the sampled points in gen-1 to gen-3 shift towards the target zone, due to identification of compounds that overcome the trade-off of the negative linear correlation between Δ-SCF gap and $r_{ND}$ present in gen-0 (Figure 2). Although only 200 complexes are sampled at each subsequent generation, we discover numerous TMCs in the target zone once their DFT properties are explicitly computed. This enrichment is greatest in gen-2, where we identify 14 new TMCs that fulfill the design criteria, leading to a rather high (i.e., 7%) lead conversion rate (i.e., number of leads over number of samples). A conservative estimate using the binomial distribution shows that one would need to sample 200,000 TMCs randomly in the base complex space to produce 14 lead complexes, demonstrating a 1,000-fold acceleration of our active learning approach compared to random sampling. In addition, the ML models improve systematically as active learning proceeds from gen-0 to gen-3, as exemplified by reduction in relative MAEs of predicting the DFT results for the set-aside test data with increasing model generation (Figure 2, Supporting Information Figure S5, see *Methods*).

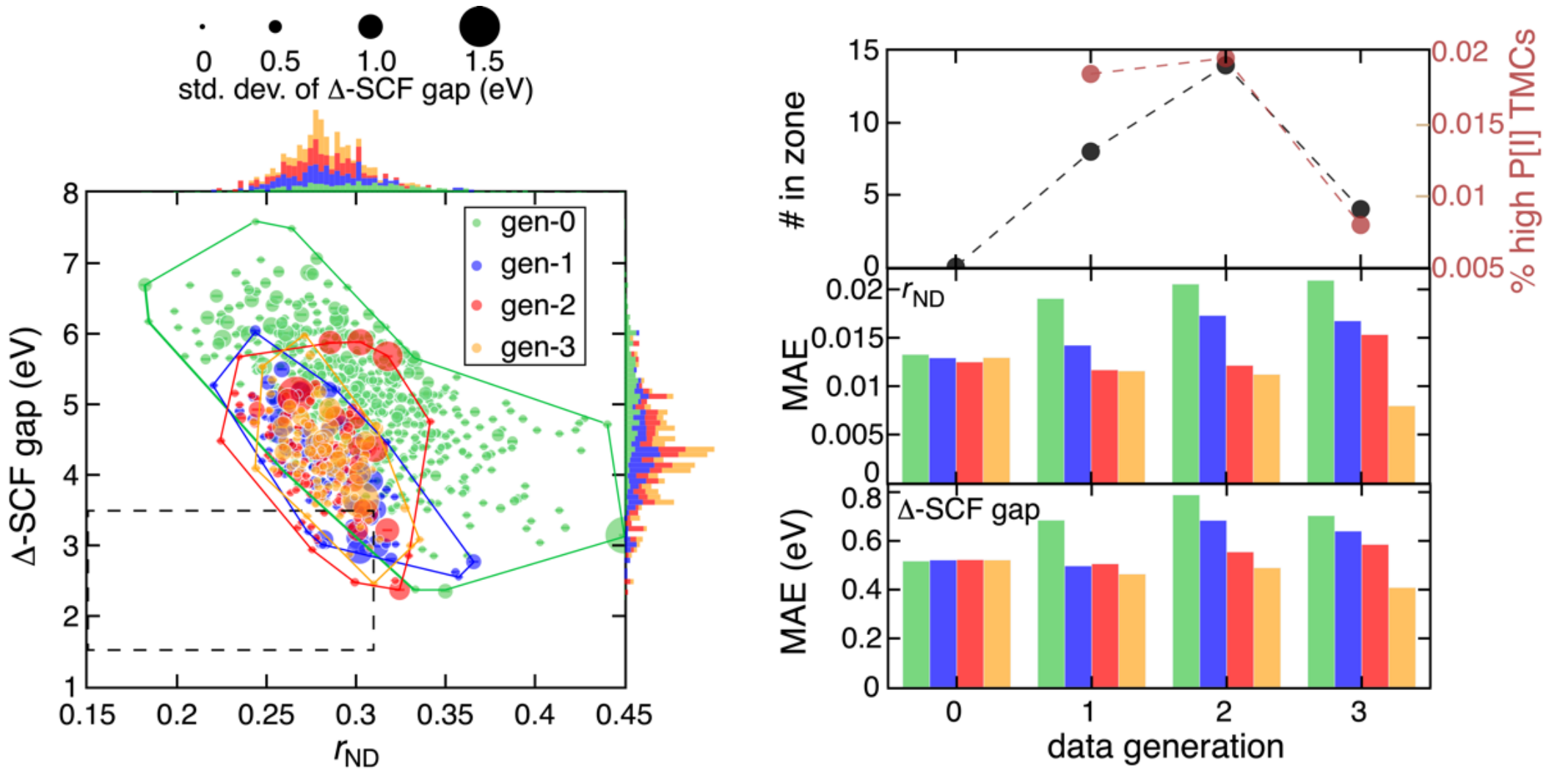

**Figure 2.** (left) DFT-computed Δ-SCF gap vs. $r_{ND}$ for base complexes in gen-0 to gen-3. For each complex, the average Δ-SCF gap over all DFAs is shown as a circle sized by the corresponding std. dev. over all DFAs. The range of values sampled in each generation is indicated by a convex hull. The target zone is shown as a rectangle with dashed lines. Normalized stacked marginal histograms for Δ-SCF gap and $r_{ND}$ are also shown. (right) The number of complexes in the target zone (black) and the percentage of the design space that has a 2D P[I] > 1/3 (brown) at each generation (top). The 2D P[I] at gen-0 is not available as the ML models have only been trained after gen-0. MAE for $r_{ND}$ (middle) and Δ-SCF gap (bottom) from gen-0 to gen-3. At each generation, the ML models are trained on the combined training set of all previous generations and are tested on the set-aside test set of each generation separately. For example, the gen-2 model (blue bars) was trained on the combined training set of gen-0, gen-1, and gen-2 data. Generations are colored as follows throughout: green for gen-0, blue for gen-1, red for gen-2, and orange for gen-3. Gen-0 represents a *k*-medoids sampling of the 1.3M base TMC space.

After three generations of active learning, both the number of TMCs landing in the target zone (i.e., after computed by DFT) and the percentage of high 2D P[I] complexes decrease, indicating that most candidate base TMCs have likely been identified as well as the leveling off of ML model performance on the 1.3M base complexes. Therefore, we used the gen-3 models to screen through the base complex design space to reveal chemical trends for the 2,432 TMCs that have a reasonable probability of residing in the target zone (i.e., 2D P[I] > 1/6). Here, we use a smaller cutoff for 2D P[I] (i.e., 1/6 compared to 1/3) to retain a reasonable number of complexes for statistical analysis. From this set, we find that complexes with Co(III) and strong-field ligands (e.g., coordinating atom combinations of: CC, CN, NP, and PP) are significantly enriched (Figure 3). This likely occurs due to our requirement of a LS ground state during the screening procedure (Figure 1). Because we prefer a small Δ-SCF gap, the complexes that are favored by 2D P[I] tend to have large ligands, consistent with our previous observation that Δ-SCF gap has a negative linear correlation with complex size[61] (Figure 3). Lastly, we find that complexes with reasonable (i.e., > 1/6) 2D P[I] tend to consist of ligands that are more saturated, as measured by their increased inverse average bond order[56] (Figure 3). This trend can be understood by the fact that unsaturated ligands tend to contain higher MR character, and complex properties correlate (i.e., are additive) with those of their constituent ligands[56] (Supporting Information Figure S6). In general, we learn from our ML models that a complex with Co(III) and large, strong-field, and relatively saturated ligands would have an increased chance of being a transition metal chromophore with the desired properties.

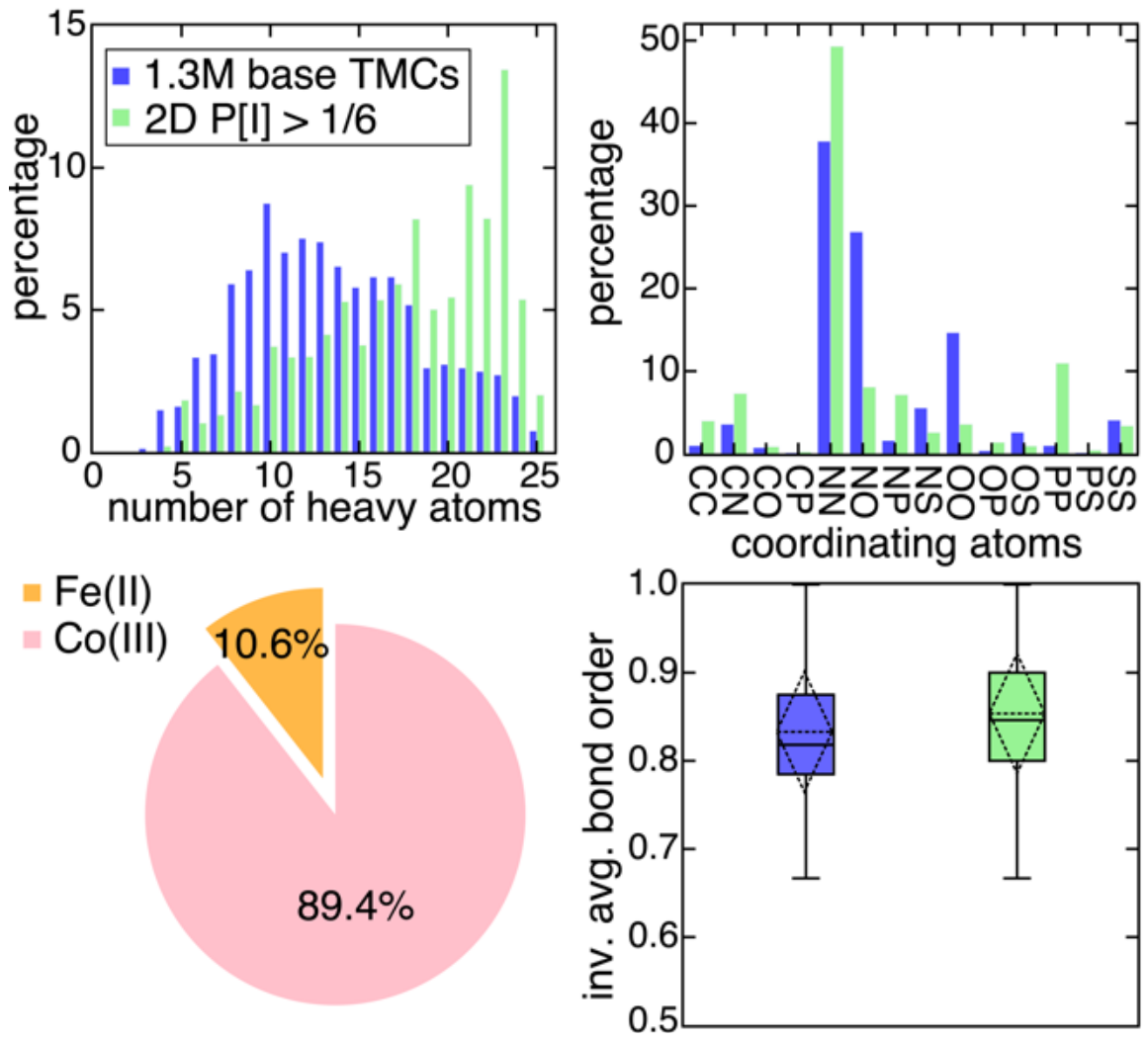

**Figure 3.** Comparison of property distributions of the 2,432 complexes with 2D PI > 1/6 evaluated by gen-3 ML models (green) and the 1.3M base complexes (blue). Bar plots for the average number of heavy atoms in the ligands involved in the 2,432 complexes (top left) and their coordinating atom types (top right), a pie chart for the core metal (orange for Fe(II) and pink for Co(III), bottom left), and a box plot for inverse average (inv. avg.) bond order for the ligands (bottom right). For each box, the median is shown as a horizontal solid line, the mean and std. dev. are shown as a dashed diamond, and the two extrema are shown by the vertical bar.

**Exploring the functionalized design space.** Hammett tuning, i.e., functionalization on conjugated rings, is a common procedure applied in experiments to fine tune the electronic properties of a TMC without dramatically scarifying its synthesizability.[64,65] We considered three possible functionalizable positions, categorized by the bond depth (*d*) of the H atom on a ring with respect to the metal (Figure 4). For a six-membered ring, *d*=3, 4, or 5 corresponds to the ortho, meta, and para positions, respectively. For a five-membered ring, only *d*=3 or 4 is feasible. We consider a wide range of 10 electron-donating or electron-withdrawing functional groups (Figure 4). To retain good likelihood of synthesizability, we constrain the *in silico* functionalization procedure to consist of one unique functionalizable position and one unique functional group for a TMC and disallow any combinations with multiple functionalizable positions or functional groups. Despite this constraint, the base design space will roughly be expanded by a factor of 25 after accounting for rings that are not functionalizable, leading to 32.5M TMCs (Figure 1). However, the effect of Hammett tuning is not expected to be large enough to move all of the base complexes into the target zone, so those far enough from the target zone can be immediately discarded. For a representative Co(III) complex that resides in the target zone, we find that Hammett tuning can roughly tune the Δ-SCF gap by 1.0 eV and $r_{\mathrm{ND}}$ by 0.04 (Figure 4). Therefore, we use the gen-3 ML models to screen through the 1.3M base complexes and only keep TMCs with a predicted Δ-SCF gap < 4.5 eV and $r_{\mathrm{ND}} < 0.35$ as candidates for Hammett tuning. These promising 30.1k base complexes lead to a design space of 0.8M functionalized TMCs for further exploration.

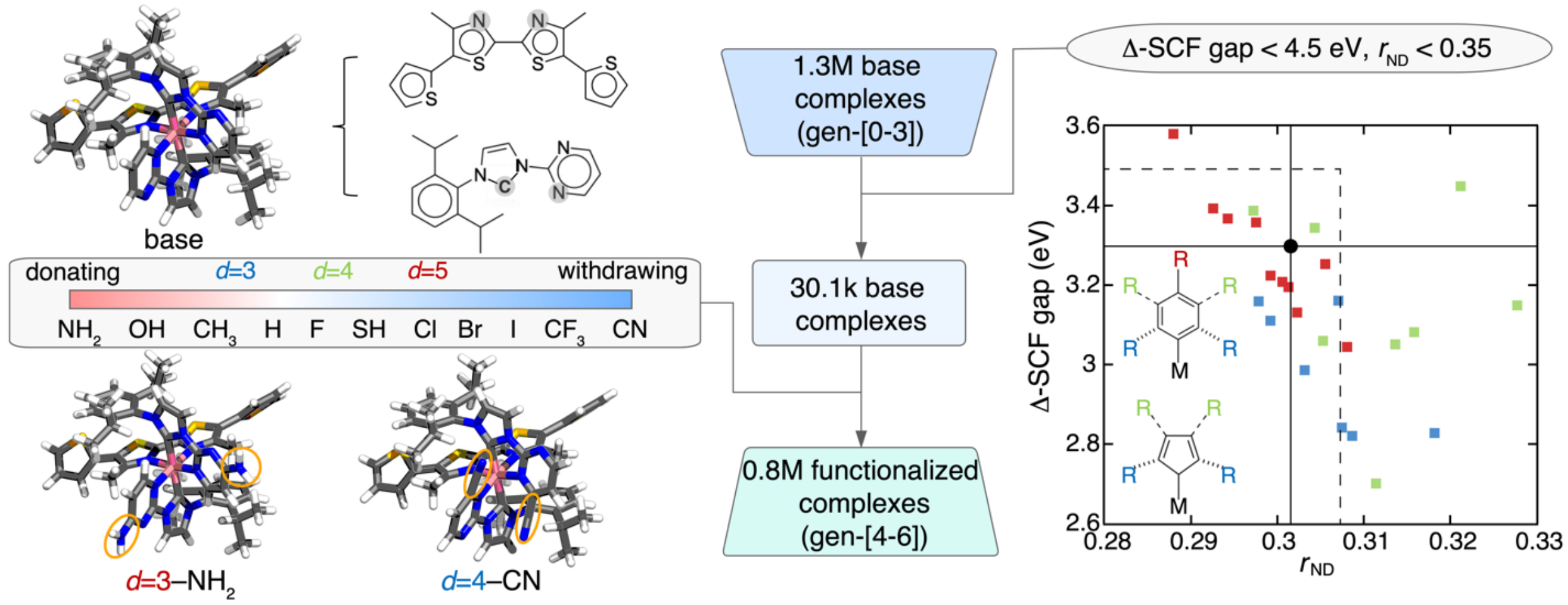

**Figure 4.** Procedure for constructing the functionalized TMC design space. (middle) The 1.3M base complexes used in gen-0 to gen-3 are first filtered down to 30.1k base complexes that are predicted to have a LS ground state, an average Δ-SCF gap < 4.5 eV, and $r_{ND}$ < 0.35, based on gen-3 ML models. These complexes are then functionalized on the coordinating rings with a chosen position (i.e., *d*=3, 4, or 5) and functional group, enlarging the design space to 0.8M functionalized TMCs to be used in gen-4 to gen-6. (left) Example of functionalizing the base complex Co(III)$(C_{19}H_{22}N_4)_2(C_{16}H_{12}N_2S_4)$. The base complex and corresponding ligands are shown at the top, where the coordinating atoms are shaded in gray. The functional groups used to perform Hammett tuning are shown in the middle. The two functionalized complexes (left with CN at the *d*=3 position and right with $NH_2$ at the *d*=4 position) are shown at the bottom. (right) Average Δ-SCF gap vs. $r_{ND}$ for functionalized Co(III)$(C_{19}H_{22}N_4)_2(C_{16}H_{12}N_2S_4)$ at each possible position (blue for *d*=3, green for *d*=4, and red for *d*=5) and functional group. The target zone is shown as a rectangle with dashed lines. The predicted properties of Co(III)$(C_{19}H_{22}N_4)_2(C_{16}H_{12}N_2S_4)$ are shown as a black circle intersected with solid lines. The insets show the rule of functionalizing a six-membered and five-membered ring, respectively.

Since we have created a new design space with functionalized complexes that our ML models have not seen before, we expected that the 2D P[I] computed based on the trained model predictions and uncertainties would not be able to directly guide the exploration of candidate chromophores. Therefore, we repeated the *k*-medoids sampling that we performed in gen-0 but this time limited to the new 0.8M functionalized design space, selecting a set of 200 complexes. Indeed, we find that the gen-3 models have significantly higher errors on the predictions of the *k*-medoids sampled gen-4 data (Figure 5). However, the ML models improve quickly after retraining on the functionalized complexes in gen-4. Therefore, we expect the 2D P[I] to regain its predictive power and undertook two generations of active learning using 2D P[I] criteria. During these two generations, the ML models achieve MAEs that are comparable to those on the base complexes (Figure 5). Because we have already isolated a promising fraction of the functionalized TMC space, both the number of TMCs landing in the target zone and the percentage of high (i.e., > 1/3) 2D P[I] complexes increase relative to the previous three generations (Figure 5). At both gen-5 and gen-6, 19 (i.e., 10%) of the sampled functionalized TMCs are verified as candidate transition metal chromophores by DFA consensus. This number greatly surpasses the average (i.e., 6) and the maximum (i.e., 14) in the previous generations. More importantly, the sampled functionalized complexes in gen-5 and gen-6 further expand the convex hull in the 2D property space, with their distributions shifted towards the target zone. We find that functionalization (i.e., Hammett tuning) indeed pushes more complexes into the target zone by fine-tuning their electronic properties. For example, the $d$=3 $CH_3$ functionalization of the base complex Co(III)($N_2C_{16}H_{12}S_4$)($N_4C_{18}H_{16}$)$_2$ (Δ-SCF gap=2.83 eV, $r_{ND}$=0.300) leads to complex **F** (Δ-SCF gap=2.39 eV, $r_{ND}$=0.305), which lowers the Δ-SCF gap with a better $r_{ND}$ value compared to other base complexes sampled in the active learning process (Figure 6, Supporting Information Table S2). This observation showcases the effectiveness of our strategy for using Hammett tuning to further enrich a pool of candidate chromophores and improve their electronic properties.

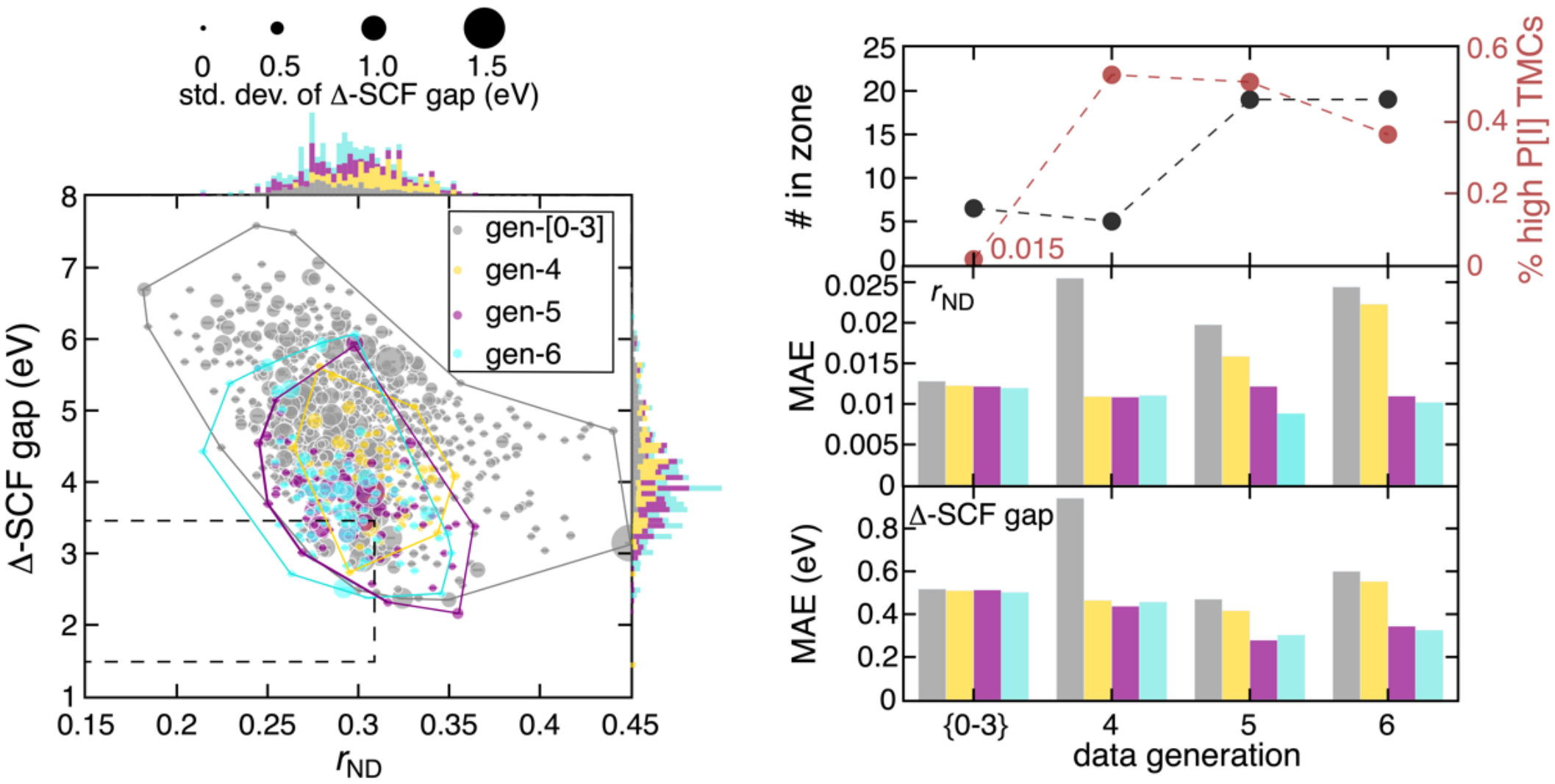

**Figure 5.** (left) DFT-computed Δ-SCF gap vs. $r_{ND}$ for functionalized complexes from gen-4 to gen-6 (yellow for gen-4, purple for gen-5, and cyan for gen-6). The base complexes in gen-0 to gen-3 are combined as gen-[0-3] (gray). For each complex, the average Δ-SCF gap over all DFAs is shown as a circle scaled by the corresponding std. dev. of Δ-SCF gaps. The range of values sampled in each generation is indicated by a convex hull. The target zone is shown as a rectangle with dashed lines. Normalized stacked marginal histograms for the Δ-SCF gap and $r_{ND}$ are also shown. (right) The number of complexes in the target zone (black) and the percentage of the design space that has a 2D P[I] > 1/3 (brown) at each generation (top), with the average shown for the combined gen-[0-3]. MAE for $r_{ND}$ (middle) and Δ-SCF gap (down) at each generation. At each generation, the ML models are trained on the combined training set of all previous generations and are tested on the set-aside test set of each generation separately. For the combined gen-[0-3], the MAEs are evaluated on the combined set-aside test sets from gen-0 to gen-3 using the gen-3 ML models. Gen-4 represents a *k*-medoids sampling of 200 TMCs on the 0.8M functionalized TMC space.

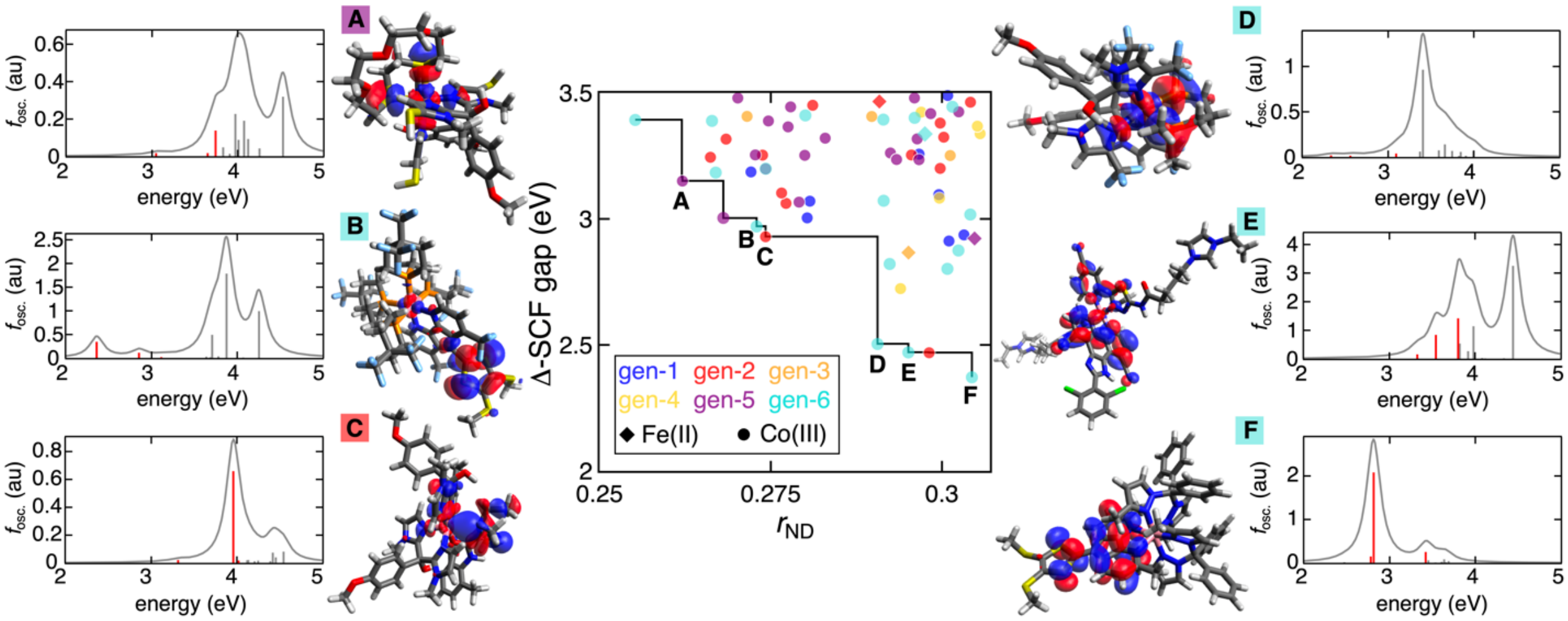

**Figure 6.** 69 TMCs sampled through the active learning process that have a LS ground state and land in the target zone computed by DFT, colored by generation and with unique symbols for each metal center (as indicated in the inset legend). The trade-off of best $r_{ND}$ values for a given Δ-SCF

gap is indicated by the black solid lines. Six out of nine TMCs are verified to have desired excited state properties (i.e., an excited state lower than 3.5 eV and MLCT character) by TDDFT calculations. These six complexes are indicated by the letters **A–F**, where the absorption spectra and the lowest energy transition orbital are shown. The lowest three absorption energies are colored red for better visibility due to the large variance among the oscillator strengths for different excitations.

To verify that complexes discovered through the active learning process have the desired excited state properties of transition metal chromophores, we computed the excited state properties with TDDFT of lead chromophores with the lowest $r_{\mathrm{ND}}$ for representative Δ-SCF gap values in the target zone from all seven generations (see *Methods*). Using this approach, we verified that six out of nine of our lead complexes have desired transition energy < 3.5 eV, where the photoexcitation process involves a MLCT state (Figure 6). This observation is surprising because we do not explicitly set design objectives that involve explicit excited state calculations of our complexes during the active learning. Still, we achieve high likelihood (i.e., 67%) for obtaining lead complexes with promising excited state properties by carefully crafting the ground-state design objectives and using DFA consensus. The remaining three complexes have excited states with energy < 3.5 eV but are "dark" states (i.e., with small oscillator strengths), which cannot be foreseen by the ground state calculations that we performed during the active learning procedure (Supporting Information Table S3). If we had not required a consensus low Δ-SCF gap (i.e., < 3.5 eV) and LS ground state, the likelihood for obtaining leads with promising excited state properties would have been much lower, at around 15% (Supporting Information Table S4).

Since all the bidentate ligands considered are synthetically accessible, we propose these complexes as promising candidates for experimental verification. The quality of the leads selected by the consensus approach is expected to be better than from a standard single-functional screen because the functionals from different rungs agree, and previous work has shown DFT consensus more frequently produces leads similar to those that are experimentally validated.[47] Moreover, although complexes **A**–**F** are not in the CSD, they all contain ligands present in other synthesized compounds that have demonstrated photo-induced properties (Supporting Information Table S5). For example, despite the fact that complex **F** has not been characterized experimentally, one of its constituent ligands (i.e., 4',5'-diaza-9'-[4,5-bis(methylthio)-1,3-dithiol-2-ylidene]fluorene) has been studied[66] for its interesting nonlinear optical properties in Co, Cu, and Cd complexes (Figure 6). These observations showcase the power of our strategy for identifying experimentally relevant candidate transition metal chromophores that were potentially missed by previous experimental exploration due to the combinatorial challenges in chemical discovery.

## Conclusions

We applied EGO with a 2D P[I] criteria to discover potential $3d^6$ Fe(II)/Co(III) transition metal chromophores in a design space of 32.5M compounds that simultaneously fulfill three design objectives: LS ground state, a Δ-SCF gap corresponding to an electronic transition in the visible region of the electromagnetic spectrum, and weak MR character (i.e., low $r_{\mathrm{ND}}$). We avoid common biases that arise from DFA choice in VHTS and ML-accelerated chemical discovery by applying a DFA-consensus approach that considers the property evaluations from 23 DFAs that span multiple rungs of "Jacob's ladder". We also addressed the challenge of synthesizability for computationally designed functional molecules by constraining the design space construction to ligands that are synthetically accessible and symmetry classes that are easy to access in experiments. Compounds discovered through this active learning workflow therefore have a higher likelihood to be synthesizable and higher fidelity (i.e., more robust to changes in DFA choice).

Despite the scarcity of potential transition metal chromophores in our design space, judged by the fact that no compounds in the initial 2,000 samples land in our target objective zone, our active learning process gradually shifts the distributions of sampled compounds towards the target zone and successfully identifies many leads. A conservative estimate suggests that our active learning approach achieves a 1,000-fold acceleration relative to the random sampling. Interrogation of our ML models revealed that Co(III) complexes with large, strong-field, and relatively saturated ligands are preferred as candidate transition metal chromophores. To fine-tune their electronic properties, we further introduced Hammett tuning by functionalization of the base complexes, which further increased the number of complexes that satisfied the design criteria. Lastly, we performed TDDFT calculations on the nine most promising leads. We found that six of the nine compounds demonstrated desired excited state properties with MLCT states and contain ligands that have been previously studied experimentally due to their interesting optical properties. We expect our strategy for design space construction and DFA-consensus enhanced active learning workflow to be broadly useful in discovering candidate molecules and materials that are more synthesizable and computationally robust in transition metal chemical space.

## Methods

**DFT calculation details.** All initial geometries were generated using molSimplify[67,68], where initial ligand geometries were derived from the crystal structures of transition metal complexes containing the ligands (Supporting Information). DFT geometry optimizations were carried out using TeraChem[69], as automated by molSimplify[67,68] with a 24 h wall time per run with up to five resubmissions. These calculations used the B3LYP[70-72] global hybrid functional with the LACVP* basis set, which corresponds to the LANL2DZ[73] effective core potential for transition metals (i.e.,

Fe, Co) and heavier elements (i.e., I or Br) and the 6-31G* basis for all remaining elements. These geometries were optimized using the L-BFGS algorithm in translation rotation internal coordinates (TRIC)[74] to the default tolerances of $4.5 \times 10^{-4}$ hartree/bohr for the maximum gradient and $10^{-6}$ hartree for the energy change between steps. All HS (i.e., quintet) states were calculated with an unrestricted formalism and LS (i.e., singlet) states with a restricted formalism. In all calculations, level shifting of 0.25 Ha was employed between the occupied and virtual spin orbitals. Geometry checks were applied to eliminate optimized structures that deviated from the expected octahedral shape following previously established metrics without modification.[75] Open-shell structures were also removed from the data set following established protocols if the expectation value of the $S^2$ operator deviated from its expected value[75] of $S(S + 1)$ by >1 $\mu_B^2$ (Supporting Information Table S6).

For optimized TMCs, we followed our established protocol[47] for the calculation of the Δ-SCF gap with multiple DFAs using a developer version of Psi4 1.4[76]. We adopted a consistent spin state convention[47]: we removed a majority-spin electron from the $N$-electron reference for the $N$-1-electron calculation and added a minority-spin electron for the $N$+1-electron case. The Δ-SCF gap is then computed as $2*E[N] - (E[N-1] + E[N+1])$. In this workflow, the converged wavefunction obtained from the B3LYP geometry optimization was used as an initial guess for the single-point energy calculations with other DFAs, thus maximizing the correspondence of the converged electronic state among all DFAs and also reducing the computational cost. We use 23 DFAs as in our previous work[47] that were chosen to be evenly distributed among the rungs of "Jacob's ladder"[77] (Supporting Information Table S7).

We evaluated the $r_{ND}$ diagnostic[57,58,78] by performing finite-temperature DFT[79] calculations using TeraChem[69]. Specifically, we followed a literature recommendation[57,58,78] to use a temperature of 9,000 K for B3LYP. Here, we evaluated fractional occupation numbers (FON) from a broadened distribution (i.e., with Fermi–Dirac statistics).

We performed linear response TDDFT calculations with the Tamm-Dancoff approximation using ωB97X-D/def2-TZVP following by the recommendation of a recent benchmark study[80] in Psi4 1.4[76]. We used a polarizable continuum implicit solvent model with water ($\varepsilon = 80$) as the solvent. Because we focus most on the lowest few excitations, only the first 30 states were computed. We broadened simulated spectra using Lorentzian functions, and we considered only excited states with significant oscillator strength (i.e., $f_{osc} > 0.01$ a.u.).

**ML models.** As in our prior work, we use extended revised autocorrelations[81,82] (eRACs) as descriptors for all our machine learning models. The eRAC features are sums of products and differences of six atom-wise heuristic properties (i.e., topology, identity, electronegativity, covalent radius, nuclear charge, and group number in the periodic table) on the 2D molecular graph. As motivated previously on large TMCs[43], we applied the maximum bond depth of four and eliminated RACs that were invariant over the mononuclear octahedral transition metal complexes. We used metal oxidation state and total ligand charge of a complex as two additional features. Because we would like to discover transition metal chromophores with a LS ground state with certain ranges of Δ-SCF gap and $r_{ND}$, we built ML models to predict these three properties. Specifically, we built i) a classification model to predict whether a complex fulfills the consensus LS condition (i.e., > 16 DFAs categorize the ground state to be LS), ii) a regression model to predict $r_{ND}$ of a LS complex, and iii) 23 separate models to predict the Δ-SCF gap of a LS complex for each DFA (Supporting Information Table S1). In our workflow, we first used the ground state classification model to filter out complexes that do not satisfy the consensus LS condition. We used the energy from both the HS and LS optimization of a complex as training data for the model to determine its ground spin state. On the contrary, only the LS calculation was used for building the ML models that predict Δ-SCF and $r_{ND}$. For the 23 Δ-SCF gap models, we adopted our established workflow[47] to fine-tune the 22 non-B3LYP models initialized by the weights of the B3LYP model to avoid randomness in the weight initialization and to increase the consistency between ANN models trained with DFT data derived from different DFAs.

During each generation of the active learning, we partitioned the data using a random 80%/20% train/test split and used 20% of the training data (i.e., 16% overall) as the validation set. As in our prior work[43], all ANN models were trained using Keras[83] with a Tensorflow[84] backend and Hyperopt[85] for hyperparameter selection for gen-0 data (Supporting Information Table S8). For all other generations, the models were only fine-tuned with a reduced learning rate (i.e., $10^{-5}$) on the combined training set of all previous generations. All ANN models were trained with the Adam optimizer up to 2,000 epochs, and dropout, batch normalization, and early stopping were applied to avoid overfitting.

**Supporting Information Statement**

Histogram for average ground state spin; Δ-SCF gap computed at different DFAs and spin states; $r_{ND}$ of optimized structures at LS and HS state; Ground state labeling with DFA consensus and AUC of ML classification models; Comparison of ligands' $r_{ND}$; Summary first three excited states for lead complexes that have "dark" states; Δ-SCF gap and $r_{ND}$ for functionalized counterpart of complex **F**; Ligands involved in CSD complexes with photo-induced properties; Summary of the filtering statistics during active learning; Summary of 23 DFAs; Range of hyperparameters sampled for ANN models; Data .csv files for all DFT-computed complexes, base complexes with high 2D P[I], and complexes reside in the target zone; Geometries for 812 ligands and DFT-optimized complexes; ML models for Δ-SCF gap and $r_{ND}$ regression and ground spin state classification.

ACKNOWLEDGMENT

This work was supported by the U.S. Department of Energy, Office of Science, Office of Advanced Scientific Computing, Office of Basic Energy Sciences, via the Scientific Discovery through Advanced Computing (SciDAC) program as well as by the Office of Naval Research under grant numbers N00014-18-1-2434 and N00014-20-1-2150. C.D. was partially supported by a seed fellowship from the Molecular Sciences Software Institute under NSF grant OAC-1547580. A.N. and D.W.K were partially supported by a National Science Foundation Graduate Research Fellowship under Grants #1122374 and Grant #1745302, respectively. H.J.K. holds a Sloan Fellowship in Chemistry, which supported this work. The authors acknowledge the MIT SuperCloud and Lincoln Laboratory Supercomputing Center for providing HPC resources that have contributed to the research results reported within this paper. The authors thank Adam H. Steeves for providing a critical reading of the manuscript.

## For Table of Contents Use Only

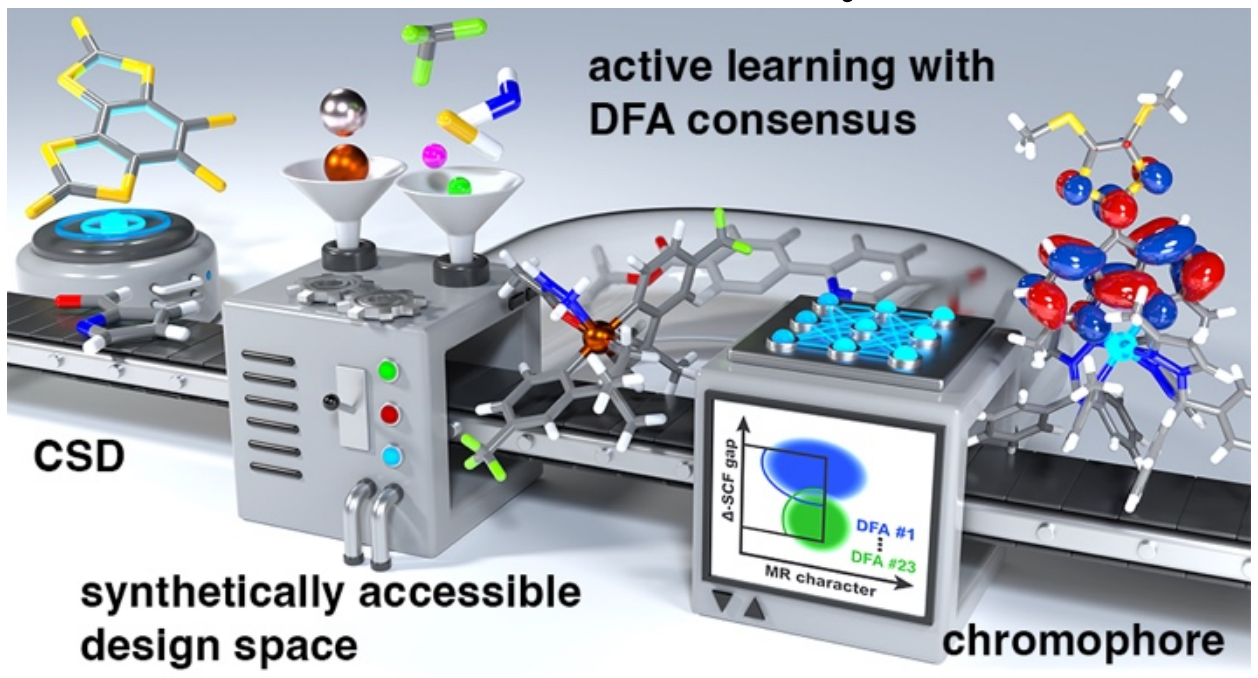

**Supporting Information for**

Active Learning Exploration of Transition Metal Complexes to Discover Method-Insensitive and Synthetically Accessible Chromophores

Chenru Duan[1,2], Aditya Nandy[1,2], Gianmarco Terrones[1], David W. Kastner[1,3], and Heather J. Kulik[1,2]

[1]Department of Chemical Engineering, Massachusetts Institute of Technology, Cambridge, MA 02139, USA
[2]Department of Chemistry, Massachusetts Institute of Technology, Cambridge, MA 02139
[3]Department of Biological Engineering, Massachusetts Institute of Technology, Cambridge, MA 02139, USA

**Contents**

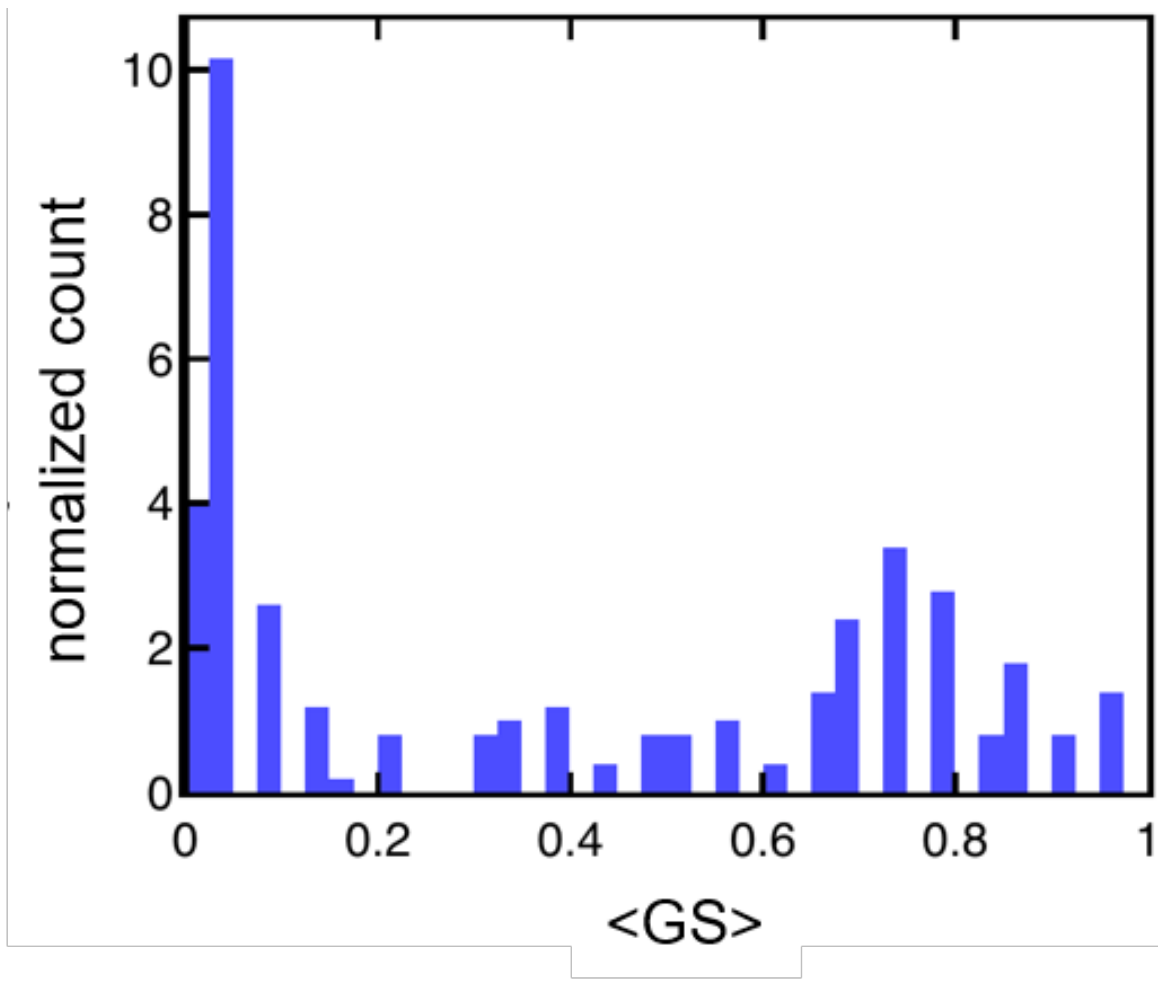

**Figure S1.** Normalized histogram for the average ground state spin (<GS>) of the gen-0 complexes, with 0 corresponding to the case where all 23 DFAs agree on the assignment of LS as the ground state and 1 corresponding to the case where all 23 DFAs agree on the assignment of HS as the ground state.

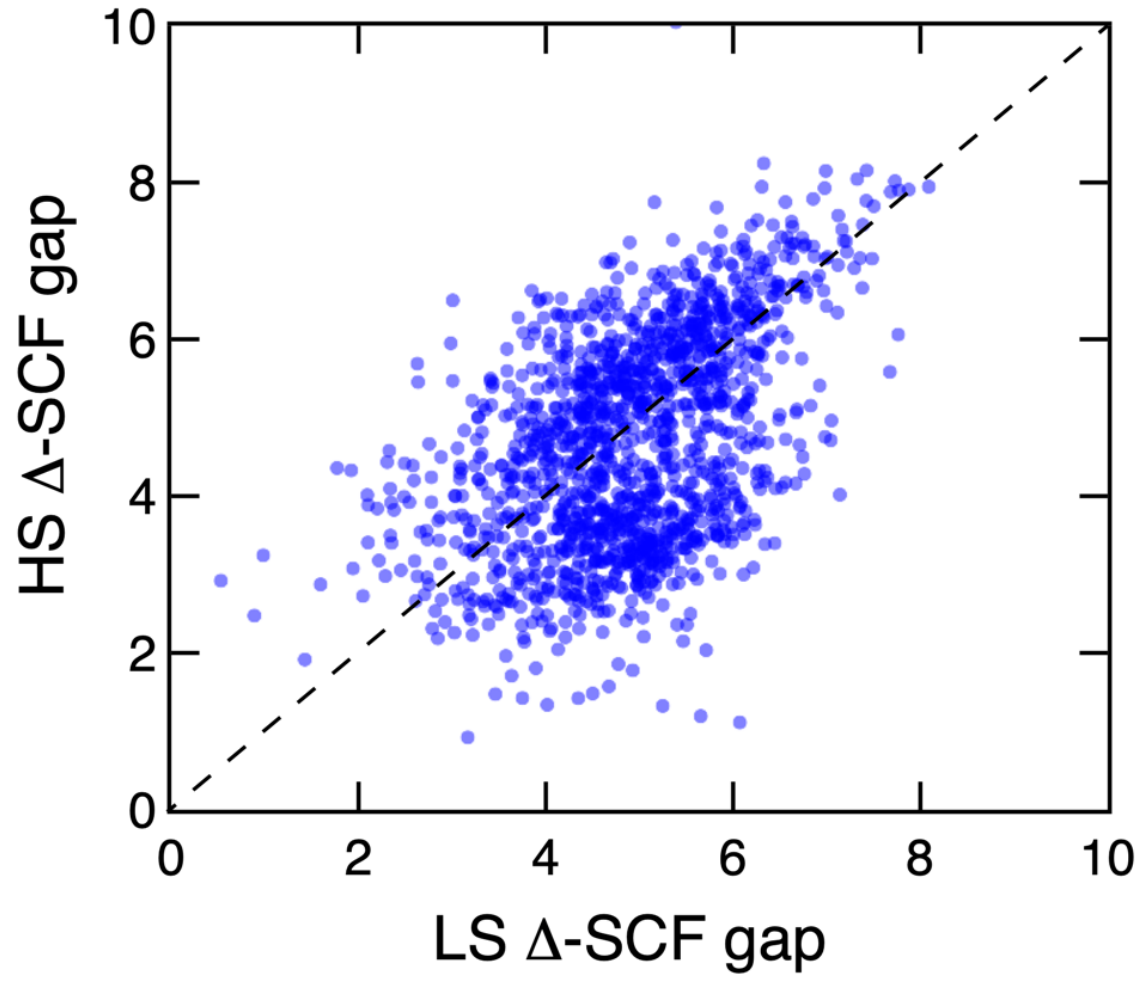

**Figure S2.** Δ-SCF gap computed at the HS versus the LS optimized geometry for the gen-0 complexes computed by B3LYP. The parity is shown as a black dashed line.

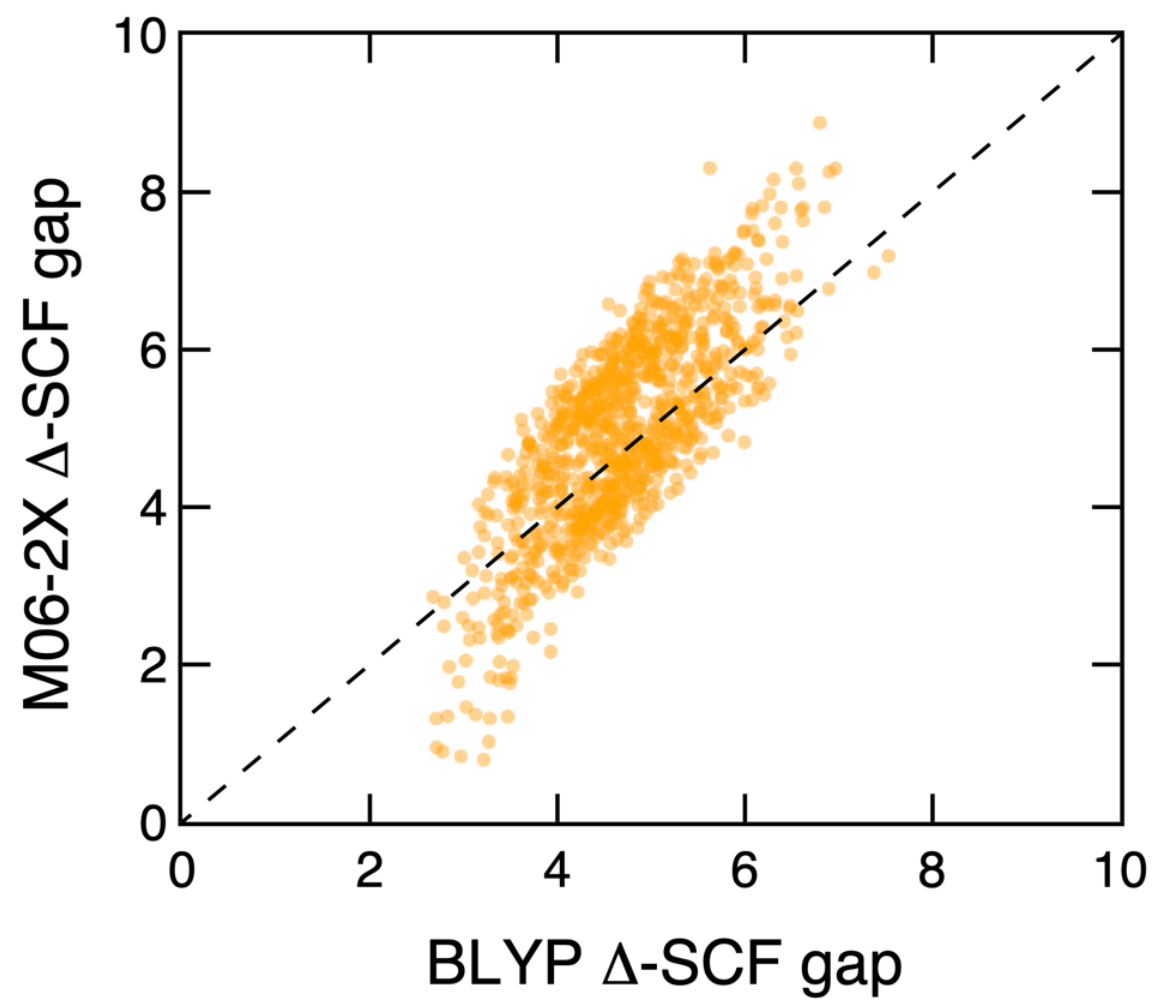

**Figure S3.** Δ-SCF gap computed at the LS optimized geometry for the gen-0 complexes computed by BLYP (*x* axis) and M06-2X (*y* axis). The parity is shown as a black dashed line.

**Table S1.** Ground state (GS) label with respect to the number and percentage of DFAs that determine LS to be the GS.

| label | number ($n$) of DFAs | percentage ($p$) of DFAs |
|---|---|---|
| consensus LS (0) | $n > 16$ | $p > 70\%$ |
| no consensus (1) | $16 \geq n > 6$ | $70\% \geq p > 30\%$ |
| consensus HS (2) | $n \leq 6$ | $p \leq 30\%$ |

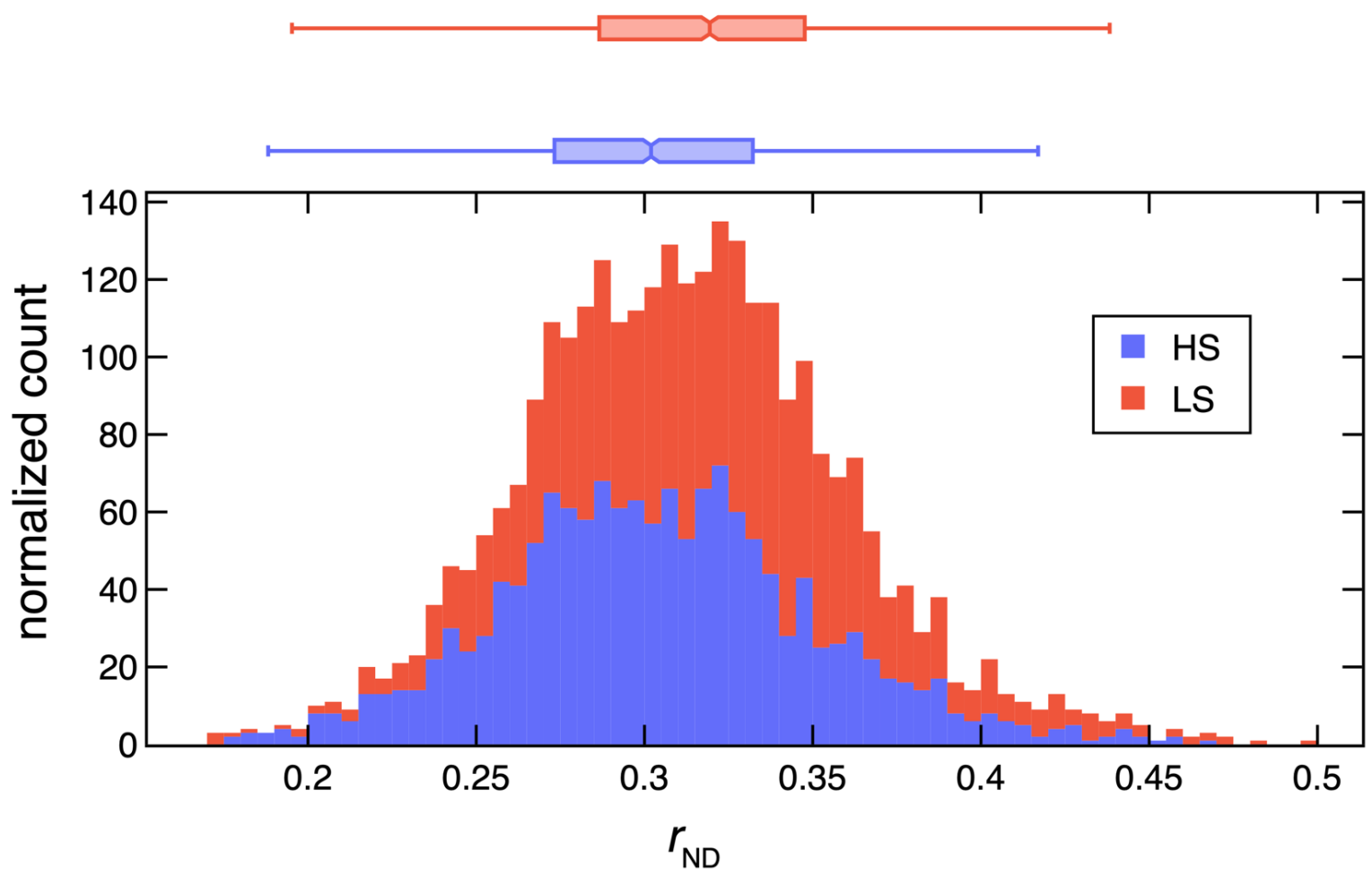

**Figure S4.** Normalized stacked histogram for $r_{ND}$ of optimized structures of the LS (red) and HS (blue) state for gen-0 complexes with box plots indicating their mean values and standard deviations at the top.

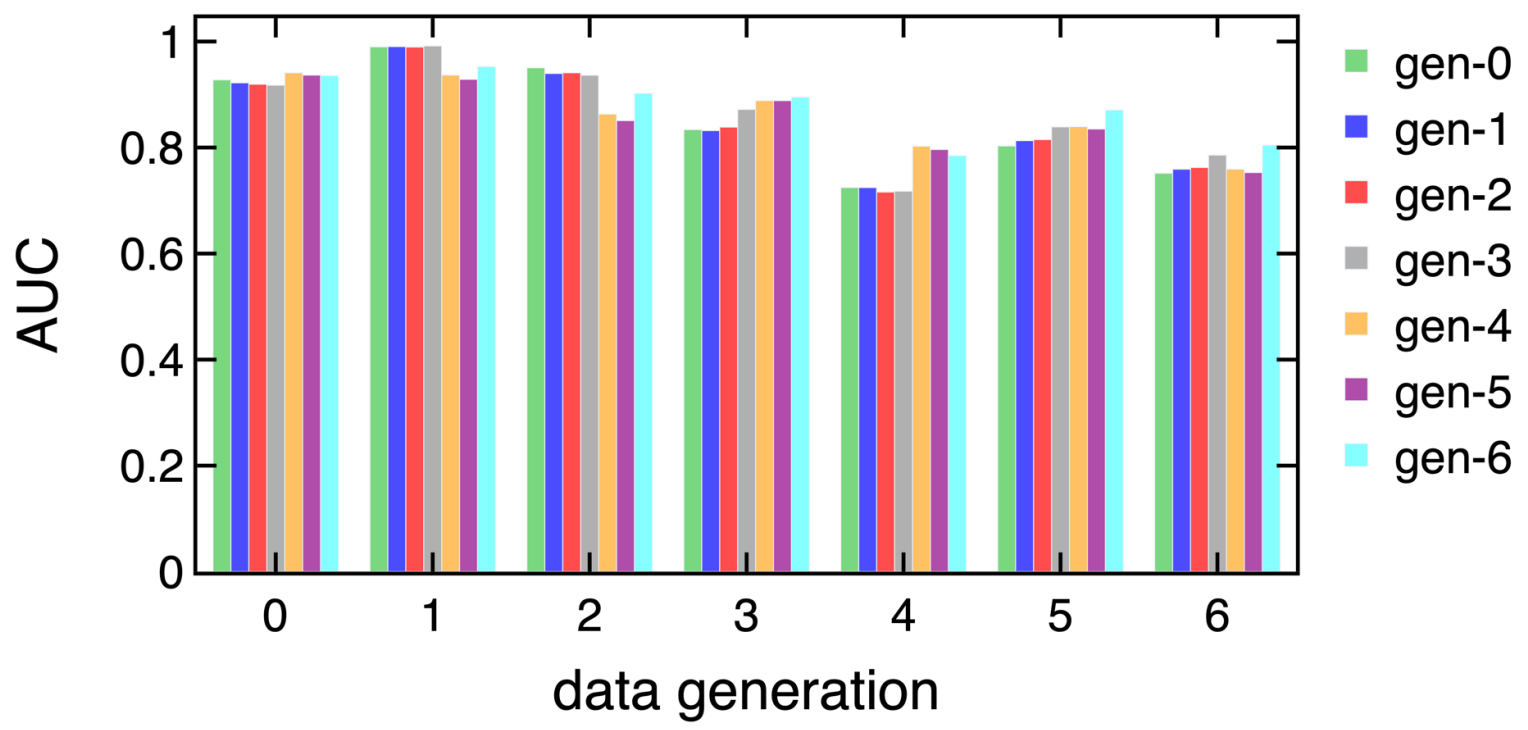

**Figure S5.** Model AUC for ground spin state classification from gen-0 to gen-6. At each generation, the ML models are trained on the combined training set of all previous generations and are tested on the set-aside test set of each generation separately. Since the number of complexes in each class and generation is small (after gen-0), the relative AUC of different models within each data generation is more meaningful than the comparisons of AUC from different data generations.

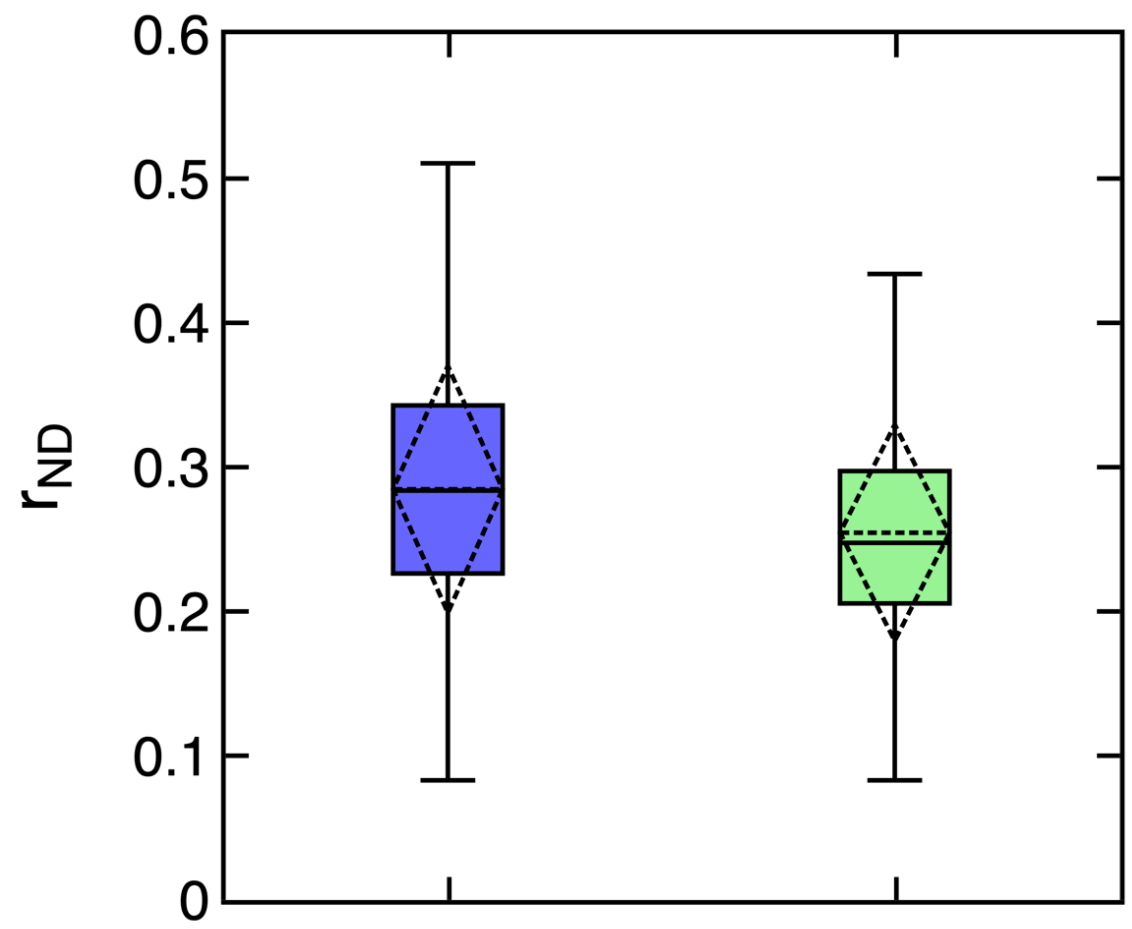

**Figure S6.** Box plot for $r_{ND}$ computed only on the ligands of the 2,432 complexes with 2D PI > 1/6 evaluated by gen-3 ML models (green) and the 1.3M base complexes (blue). For each box, the median is shown as a horizontal solid line, the mean and std. dev. are shown as a dashed diamond, and the two extrema are shown by the vertical bar. Note that the $r_{ND}$ is computed on ligands instead of transition metal complexes. Due to the ligand additivity effect, the model favors low-$r_{ND}$ ligands although the model has only seen the whole complexes.

**Table S2.** Summary of Δ-SCF gap and $r_{ND}$ for the other functionalized counterpart of complex **F** in the main text. The use of "--" corresponds to the case with no functionalization. Complex F is shaded in blue.

| depth (*d*) | functional group | I Δ-SCF gap I (eV) | $r_{ND}$ |
|---|---|---|---|
| -- | -- | 2.82 | 0.301 |
| 3 | $CH_3$ | 2.38 | 0.306 |
| 4 | $CH_3$ | 3.04 | 0.288 |
| 4 | $NH_2$ | 3.17 | 0.309 |
| 5 | $Cl^-$ | 3.14 | 0.301 |
| 5 | $F^-$ | 2.81 | 0.302 |

**Table S3.** Summary of the properties for the first three excited states for the three complexes that lie on the Pareto front but do not have desired excited state properties.

| I Δ-SCF gap I (eV) | $r_{ND}$ | 1st excited state | | 2nd excited state | | 3rd excited state | |
|---|---|---|---|---|---|---|---|
| | | energy (eV) | $f_{osc}$ (a.u.) | energy (eV) | $f_{osc}$ (a.u.) | energy (eV) | $f_{osc}$ (a.u.) |
| 3.40 | 0.256 | 2.50 | 0.001 | 2.57 | 0.001 | 2.64 | 0.000 |
| 2.48 | 0.299 | 2.43 | 0.001 | 2.48 | 0.002 | 2.56 | 0.002 |
| 3.01 | 0.269 | 2.28 | 0.000 | 2.50 | 0.000 | 2.51 | 0.000 |

**Table S4.** Summary of 13 randomly chosen complexes in gen-0 that fulfill the Δ-SCF gap requirement (i.e., < 3.5 eV) by certain DFAs but violate either the consensus LS ground state or consensus low Δ-SCF gap requirement. The ligand ID is our internal database ID, for which the corresponding geometry can be found in the Supporting Information Data. The two complexes that have promising excited state properties (i.e. with both bright state transition energy < 3.5 eV and MLCT character) are shaded in blue.

| metal | Ligand ID and count | l Δ-SCF gap l (eV) | Lowest Δ-SCF gap (eV) and the corresponding DFA | Consensus Δ-SCF gap | Consensus LS ground state | 1st bright state energy (eV) and its $f_{osc}$ (a.u.) | MLCT state |
|---|---|---|---|---|---|---|---|
| Fe | 5fad526535c34d07 3da2144e (1) 5fad590e35c34d07 3da2c0cc (2) | 2.51 | 1.61 (MN15) | Yes | No | 3.88 (0.043) | Yes |
| Co | 5fad527a35c34d07 3da22278 (2) 5fad5f8135c34d07 3da30b47 (1) | 3.69 | 2.99 (M06-2X) | No | Yes | 3.52 (0.518) | No |
| Co | 5fad579235c34d07 3da2ab70 (1) 5fad678135c34d07 3da3510d (2) | 3.92 | 3.42 (M06) | No | Yes | 3.58 (0.458) | Yes |
| Fe | 5fad567435c34d07 3da2994c (2) 5fad5b1035c34d07 3da2da94 (1) | 4.29 | 3.38 (MN15) | No | Yes | 2.71 (0.153) | Yes |
| Co | 5fad525f35c34d07 3da20be8 (1) 5fad542a35c34d07 3da26ae3 (2) | 3.83 | 3.44 (SCAN0) | No | Yes | 4.07 (0.044) | Yes |
| Co | 5fad55d135c34d07 3da28e00 (1) 5fad52a935c34d07 3da23367 (2) | 3.82 | 3.43 (MN15-L) | No | Yes | 4.11 (0.780) | No |
| Fe | 5fad526235c34d07 3da210de (2) 5fad52ca35c34d07 3da23b96 (1) | 5.06 | 3.23 (MN15) | No | No | 3.35 (0.092) | No |
| Fe | 5fad526b35c34d07 3da219c0 (2) 5fad52ac35c34d07 3da23452 (1) | 4.68 | 3.46 (MN15) | No | No | 4.78 (0.578) | No |
| Fe | 5fad54ac35c34d07 3da2770 (1) 5fad591135c34d07 3da2c0f2 (2) | 3.61 | 3.36 (PBE0) | No | No | 3.80 (0.043) | Yes |
| Fe | 5fad527935c34d07 3da22206 (2) 5fad57da35c34d07 3da2af9c (1) | 4.79 | 3.46 (MN15) | No | No | 3.93 (0.272) | No |
| Fe | 5fad62ee35c34d07 3da32b44 (2) 5fad55d135c34d07 3da28e00 (1) | 4.36 | 3.15 (MN15) | No | No | 3.75 (0.096) | Yes |
| Co | 5fad528335c34d07 3da22689 (2) 5fad565d35c34d07 3da297c6 (1) | 3.84 | 2.91 (SCAN0) | No | No | 3.22 (0.220) | Yes |
| Co | 5fad5af735c34d07 3da2d967 (2) 5fad5e5835c34d07 3da2ffa0 (1) | 3.23 | 2.94 (B3PW91) | Yes | No | 3.70 (0.150) | Yes |

**Table S5.** Summary of ligands in complexes **A**–**F** that have parent complexes in CSD that demonstrate interesting photo-induced properties and the associated references. The complex ID corresponds to the index of the six complexes in Figure 6 of the main text. The ligand ID is our internal database ID, for which the corresponding geometry can be found in the Supporting Information Data. For each ligand, the refcode for the CSD complex that contains this ligand and demonstrates interesting photo-induced properties is also shown.

| complex ID | ligand ID | ligand chemical name | CSD refcode | relevant property and reference |
|---|---|---|---|---|
| **F** | 5fad527635c34d073da22057 | 4',5'-diaza-9'-[4,5-bis(methylthio)-1,3-dithiol-2-ylidene]fluorene | MIVSIU | nonlinear optical property[1] |
| **E** | 5fad60c035c34d073da31750 | 2-(2,6-dichlorophenyl)-1H-imidazo[4,5-f][1,10]phenanthroline | ROVSEC | peak of UV-vis spectra in the visible light region[2] |
| **D** | 5fad528035c34d073da22500 | 1,2-bis(dimethylphosphino)ethane | JUWKOD | luminenscence[3] |
| **A, C, D** | 5fad542a35c34d073da26ae3 | Bis(1-methylimidazol-2-yl)-(4-methoxyphen-1-yl)methanol | EWEWAE | peak of UV-vis spectra in the visible light region[4] |
| **B** | 5fad67af35c34d073da3526f | 1,2-bis(phosphorinan-1-yl)ethane | XAZZED | photochemically induced oxidation[5] |

**Table S6.** Summary of the filtering statistics for the TMCs at each generation, including the number attempted, number of complexes with valid initial geometry (i.e., no ligand clashing), and the number of complexes with good geometry and $<S^2>$ (judged by our established protocols[6-7]) after converged geometry optimizations. Generations that explored the base complexes space are shaded in blue and those that explored the functionalized complexes are shaded in green.

| generation | attempted | valid initial geometry | converged, good geometry and $<S^2>$ |
|---|---|---|---|
| 0 | 2000 | 1822 | 1470 |
| 1 | 200 | 188 | 146 |
| 2 | 200 | 195 | 157 |
| 3 | 200 | 188 | 146 |
| 4 | 200 | 143 | 91 |
| 5 | 200 | 184 | 151 |
| 6 | 200 | 188 | 150 |

**Table S7**. Summary of 23 DFAs in the original work of Duan *et al.*[8], including the rungs on "Jacob's ladder" of DFT, HF exchange fraction, LRC range-separation parameter (bohr$^{-1}$), MP2 correlation fraction, and whether empirical (i.e., D3) dispersion correction is included.

| DFA | type | exchange type | HF exchange percentage | LRC RS parameter (bohr$^{-1}$) | MP2 correlation | D3 dispersion |
|---|---|---|---|---|---|---|
| BP86[9-10] | GGA | GGA | -- | -- | -- | no |
| BLYP[11-12] | GGA | GGA | -- | -- | -- | no |
| PBE[13] | GGA | GGA | -- | -- | -- | no |
| TPSS[14] | meta-GGA | meta-GGA | -- | -- | -- | no |
| SCAN[15] | meta-GGA | meta-GGA | -- | -- | -- | no |
| M06-L[16] | meta-GGA | meta-GGA | -- | -- | -- | no |
| MN15-L[17] | meta-GGA | meta-GGA | -- | -- | -- | no |
| B3LYP[18-20] | GGA hybrid | GGA | 0.200 | -- | -- | no |
| B3P86[9, 18] | GGA hybrid | GGA | 0.200 | -- | -- | no |
| B3PW91[18, 21] | GGA hybrid | GGA | 0.200 | -- | -- | no |
| PBE0[22] | GGA hybrid | GGA | 0.250 | -- | -- | no |
| ωB97X[23] | RS hybrid | GGA | 0.158 | 0.300 | -- | no |
| LRC-ωPBEh[24] | RS hybrid | GGA | 0.200 | 0.200 | -- | no |
| TPSSh[14] | meta-GGA hybrid | meta-GGA | 0.100 | -- | -- | no |
| SCAN0[25] | meta-GGA hybrid | meta-GGA | 0.250 | -- | -- | no |
| M06[26] | meta-GGA hybrid | meta-GGA | 0.270 | -- | -- | no |
| M06-2X[26] | meta-GGA hybrid | meta-GGA | 0.540 | -- | -- | no |
| MN15[27] | meta-GGA hybrid | meta-GGA | 0.440 | -- | -- | no |
| B2GP-PLYP[28] | double hybrid | GGA | 0.650 | -- | 0.360 | no |
| PBE0-DH[29] | double hybrid | GGA | 0.500 | -- | 0.125 | no |
| DSD-BLYP-D3BJ[30] | double hybrid | GGA | 0.710 | -- | 1.000 | yes |
| DSD-PBEB95-D3BJ[30] | double hybrid | GGA | 0.660 | -- | 1.000 | yes |
| DSD-PBEP6-D3BJ[30] | double hybrid | GGA | 0.690 | -- | 1.000 | yes |

**Table S8**. Range of hyperparameters sampled for ANN models trained from scratch with Hyperopt[31]. The lists in the architecture row can refer to one, two, or three hidden layers (i.e., the number of items in the list), and the number of nodes in each layer are given as elements of the list. The built-in Tree of Parzen Estimator algorithm in Hyperopt was used for the hyperparameter selection process.

| | |
|---|---|
| Architecture | [[128], [256], [512], [128, 128], [256, 256], [512, 512], [128, 128, 128], [256, 256, 256], [512, 512, 512]] |
| L2 regularization | [1e-6, 1] |
| Dropout rate | [0, 0.5] |
| Learning rate | [1e-6, 1e-3] |
| Beta1 | [0.75, 0.99] |
| Batch size | [16, 32, 64, 128, 256, 512] |